\documentclass[journal,comsoc]{IEEEtran}

\ifCLASSINFOpdf

\else

\fi

\usepackage{mathrsfs}
\usepackage{amsmath}
\usepackage{amsfonts}
\usepackage{amsthm}
\usepackage{amssymb}
\usepackage{graphicx}
\usepackage{subfigure}
\usepackage{indentfirst}
\usepackage{array}
\usepackage{epstopdf}
\usepackage{cite}
\usepackage{enumerate}
\usepackage{bm}
\usepackage{dsfont}
\usepackage[linesnumbered,ruled,vlined]{algorithm2e}
\usepackage{multirow}
\usepackage{verbatim}
\usepackage{stfloats}
\usepackage{color}

\SetKwComment{Comment}{//}{}
\SetKwBlock{Begin}{Function}{end}

\begin{document}

\title{Communication Beyond Transmitting Bits: Semantics-Guided Source and Channel Coding}

\author{Jincheng Dai, \IEEEmembership{Member, IEEE},
Ping Zhang, \IEEEmembership{Fellow, IEEE},
Kai Niu, \IEEEmembership{Member, IEEE},
Sixian Wang, \IEEEmembership{Graduate Student Member, IEEE},
Zhongwei Si, \IEEEmembership{Member, IEEE},
and Xiaoqi Qin, \IEEEmembership{Member, IEEE}

\thanks{\emph{Corresponding authors: Jincheng Dai, Ping Zhang}}


\thanks{The authors are with Beijing University of Posts and Telecommunications, Beijing 100876, China.}

\vspace{-1em}
}

\maketitle

\begin{abstract}
Classical communication paradigms focus on accurately transmitting bits over a noisy channel, and Shannon theory provides a fundamental theoretical limit on the rate of reliable communications. In this approach, bits are treated equally, and
the communication system is oblivious to what meaning these bits convey or how they would be used. Future communications towards intelligence and conciseness will predictably play a dominant role, and the proliferation of connected intelligent agents requires a radical rethinking of coded transmission paradigm to support the new communication morphology on the horizon. The recent concept of ``semantic communications'' offers a promising research direction. Injecting semantic guidance into the coded transmission design to achieve semantics-aware communications shows great potential for further breakthrough in effectiveness and reliability. This article sheds light on semantics-guided source and channel coding as a transmission paradigm of semantic communications, which exploits both data semantics diversity and wireless channel diversity together to boost the whole system performance. We present the general system architecture and key techniques, and indicate some open issues on this topic.
\end{abstract}

\IEEEpeerreviewmaketitle

\section{Why We Need Semantic Coded Transmission?}\label{section_introduction}

Since the masterpiece of Shannon was published in 1948 \cite{shannon1948}, the channel capacity formula has been serving as the guidance of communication system design for more than seven decades. Following that, the traditional communication system design philosophy upgrades the transmission capability mainly by stacking more spectrum resources, stronger error-correction channel codes, higher-order modulation, larger-scale antennas, and denser access points with ever-increasing complexity. In a nutshell, traditional systems are evolving under the \emph{outward-expansion} mode. While this approach has successfully served content delivery oriented wireless networks for decades, it yet results in severe high-frequency coverage costs, complicated signal processing, high energy consumption, etc. Together with extraordinary promises, naively increasing channel capacity cannot address all communication problems.

As Weaver summarized in his landmark work \cite{weaver}, communications problems in a broad sense comprise three serially escalating levels as following:

\begin{itemize}
  \item LEVEL A. The technical problem: How accurately can the symbols of communication be transmitted?

  \item LEVEL B. The semantic problem: How precisely do the transmitted symbols convey the desired meaning?

  \item LEVEL C. The effectiveness problem: How effectively does the received meaning affect conduct in the desired way?
\end{itemize}

Traditional communication systems focus on accurately transmitting bits over a noisy communication channel. Shannon theory provides a fundamental theoretical limit on the rate of reliable communications, where bits are treated equally, and the communication system is oblivious to what meaning the bits convey or how they would be used. Therefore, according to Weaver's statements, traditional systems are still focusing merely on solving the technical problem. Nevertheless, many emerging applications, from autonomous driving to Internet of everything, will involve connecting intelligent machines, where the goal is often not to reconstruct the underlying bit message, but to enable the receiver to make the right decision at the right time. Similarly, human-machine interactions will be an important component, where humans simultaneously interact with multiple devices via multimodal commands. An interesting common characteristic of all these future communication technology breakthroughs is that they attempt to introduce native intelligence as integral part of wireless networks, that drives a hierarchical upgradation from the technical level to the semantic level. This fact necessitates a radical rethinking of the communication system design that should shift from the traditional outward-expansion mode to a new \emph{inward-discovery} mode.

\begin{figure*}[t]
\setlength{\abovecaptionskip}{0.cm}
\setlength{\belowcaptionskip}{-0.cm}
  \centering{\includegraphics[scale=0.4]{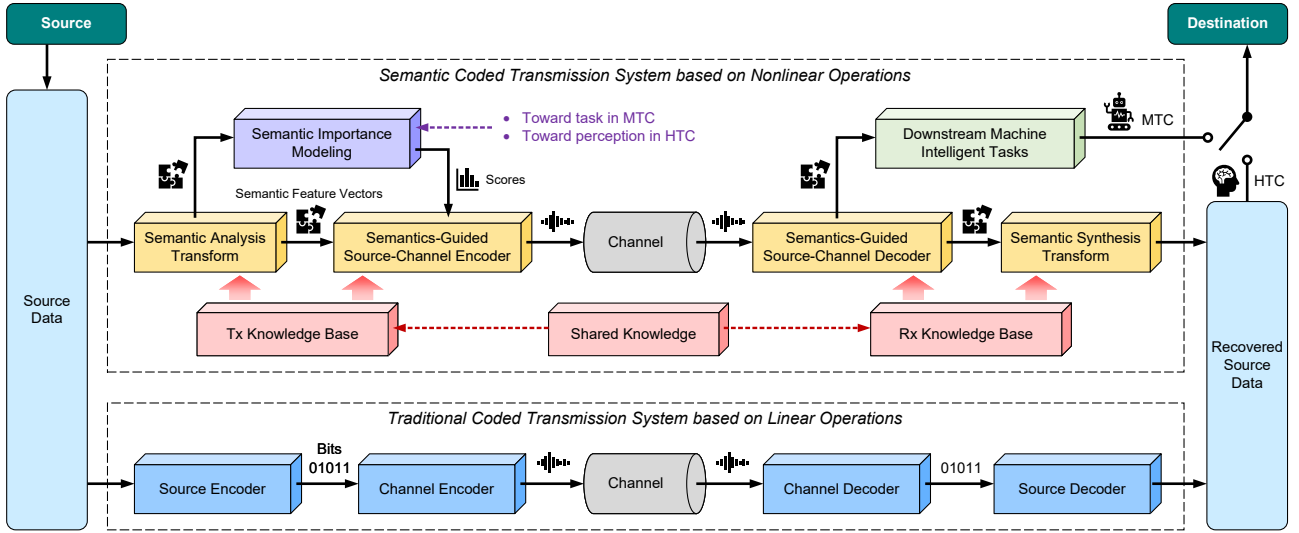}}
  \caption{Schematic diagram of a semantic coded transmission (SCT) system. Compared to classical coded transmission system with linear operations, modules in our SCT system are of nonlinear operations.}\label{Fig1}
  \vspace{-0.5em}
\end{figure*}

To this end, we investigate a new communication paradigm focusing on semantic and goal-oriented aspects, which help the transceiver identify the most valuable information, i.e., the information necessary to recover the meaning intended by the transmitter. Performance assessment goes beyond the common Shannon paradigm of guaranteeing the correct reception of each single transmitted bit, irrespective of the meaning conveyed or the goal to be achieved. A critical problem in this new paradigm is how to extract, process, and transmit semantic information. One widely-recognized solution is bridging the two branches of Shannon theory, source and channel, together to enable an end-to-end coded transmission system. The channel transmission can well match with the source semantic features, which allocates more resources to those information critical for preserving the source semantics. The source semantic feature extraction, compression, and synthesis also take into account the impacts of imperfect wireless channel transmission, e.g., noise, fading, interference, etc.

In this article, we introduce a novel unified framework of \emph{semantics-guided source and channel coding}. The corresponding end-to-end communication system can be collected under the name semantic coded transmission (SCT). A special note is that although there are many works about semantic information processing in natural language processing (NLP) and computer vision (CV) communities, they focus merely on the source representation or compression \cite{young2018recent,jing2020self}, which are not relevant to wireless communication problems. In a nutshell, traditional source semantic coding and channel coded transmission are two separated domains, where channel transmission is defaulted as error-free in source coding research, also, channel coding does not consider the source semantics. This article bridges these two domains together to enable the semantic guidance injected into joint source-channel coding design. The resulting SCT exploits \emph{data semantics diversity} and \emph{channel diversity} together to boost the whole system performance. Under this paradigm, channel transmission will be of higher distortion-tolerance capability over severe wireless channels, meanwhile, source semantic information processing will also be of better robustness.

Next, we present a brief tutorial on the general architecture and key techniques of semantics-guided source and channel coding, and outline theoretical and technical challenges and future research issues.

\section{System Architecture}

\begin{figure*}[t]
\setlength{\abovecaptionskip}{0.cm}
\setlength{\belowcaptionskip}{-0.cm}
  \centering{\includegraphics[scale=0.4]{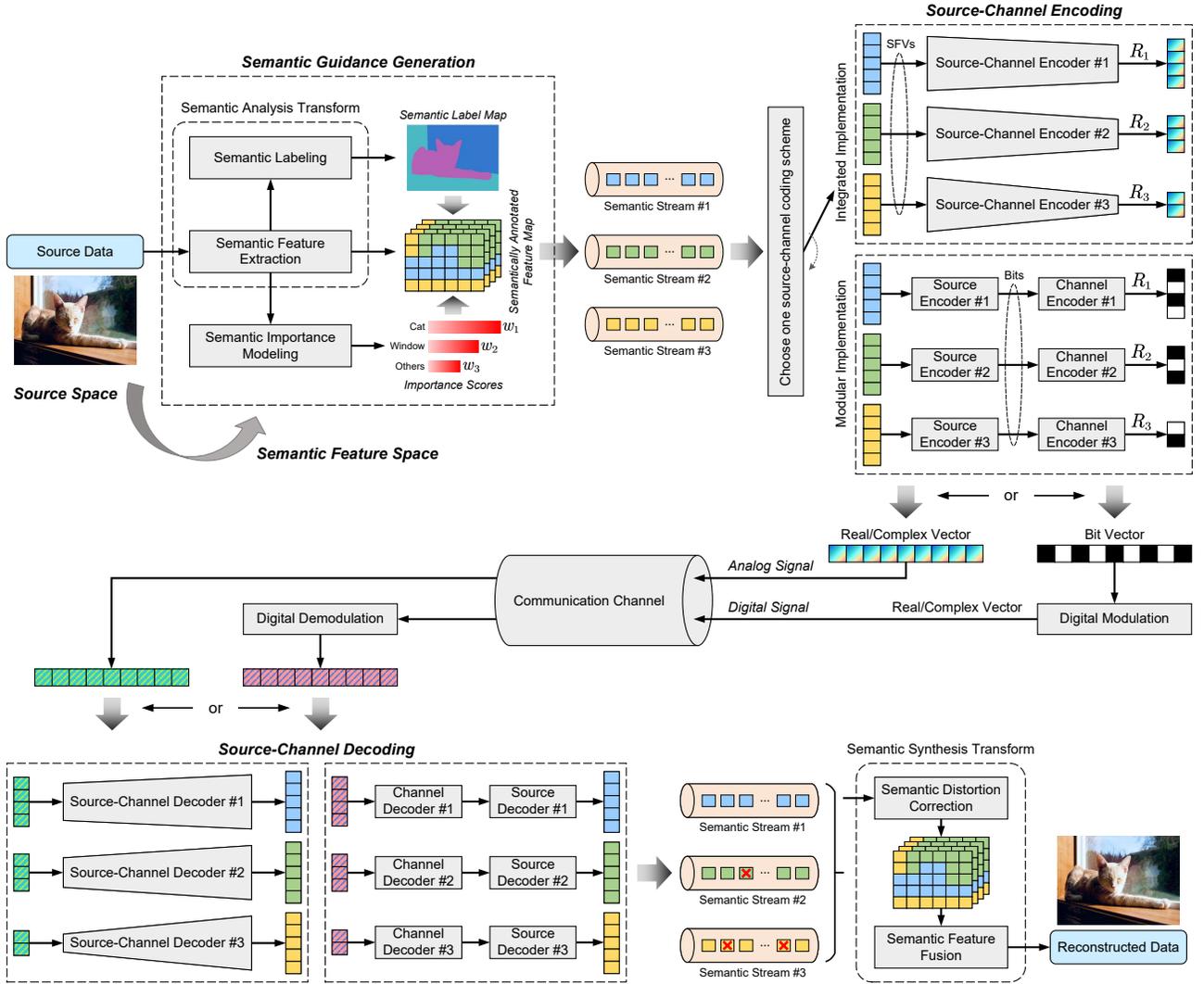}}
  \caption{The general architecture of SCT system, includes common semantic guidance generation and two types of source-channel coding implementations.}\label{Fig2}
  \vspace{0em}
\end{figure*}

By an SCT system we will mean a system of the type indicated schematically in Fig. \ref{Fig1}. It consists of essentially the following parts:


\begin{enumerate}[1.]
	\item A \emph{source} that produces data to be transmitted to the receiver. Source data may be of various types, including text, speech or audio, image, video, and advanced visual media, e.g., point clouds, volumetric video, etc. The source can also produce various combinations above types, for example, multimedia data containing video and an associated audio.
	
	\item A \emph{semantic analysis transform} module that operates on the source data to extract its semantic features and produce semantically annotated messages. The resulting semantically annotated feature map is segmented as multiple semantic streams, each contains a semantic feature vector (SFV) whose elements belong to the same semantic object.

    \item A \emph{semantic importance modeling} module that assesses the semantic value of each SFV. This function is highly tied with the communication purpose in different scenarios, e.g., the human-type communication (HTC) or the machine-type communication (MTC).
	
	\item A \emph{source-channel encoder} that operates on every SFV in some way to produce signals suitable for transmission over the channel. Specific designs of this module are guided by the semantic importance scores. The compression (source coding) and error correction (channel coding) of this encoder can be realized either in a modular way as that in traditional systems, or in an integrated way that follows the joint source-channel coding theory \cite{fresia2010joint}.
	
	\item The \emph{channel} is merely the physical medium used to transmit the signal from transmitter to receiver, which is identical with that in the classical systems.
	
	\item The \emph{source-channel decoder} performs the inverse operation of that done by the source-channel encoder, reconstructing the SFVs.
	
	\item The \emph{semantic synthesis transform} module also performs the inverse operation of that done by the semantic analysis transform. After that, the semantic feature fusion is performed to reconstruct the source data or directly drive the downstream machine intelligent tasks.
	
	\item The \emph{destination} is the person (or thing) for whom the source data is intended.
	
	\item The \emph{local knowledge} is implicitly given at both transmitter and receiver ends which provides \emph{a priori} information. This module might be not necessarily an entity, for example, it can also be the dataset used for training the deep neural networks that are employed for semantic guidance generation and source-channel coding such that the resulting network parameters indeed become one format of the knowledge.
\end{enumerate}

\section{Key Methodologies}

The general architecture of our SCT system is illustrated in Fig. \ref{Fig2}. Three key techniques, including semantic guidance generation, semantics-guided source-channel coding, and semantic distortion correction, will be elaborately introduced in this part.

\begin{figure*}[t]
	\setlength{\abovecaptionskip}{0.cm}
	\setlength{\belowcaptionskip}{-0.cm}
	\centering{\includegraphics[scale=0.35]{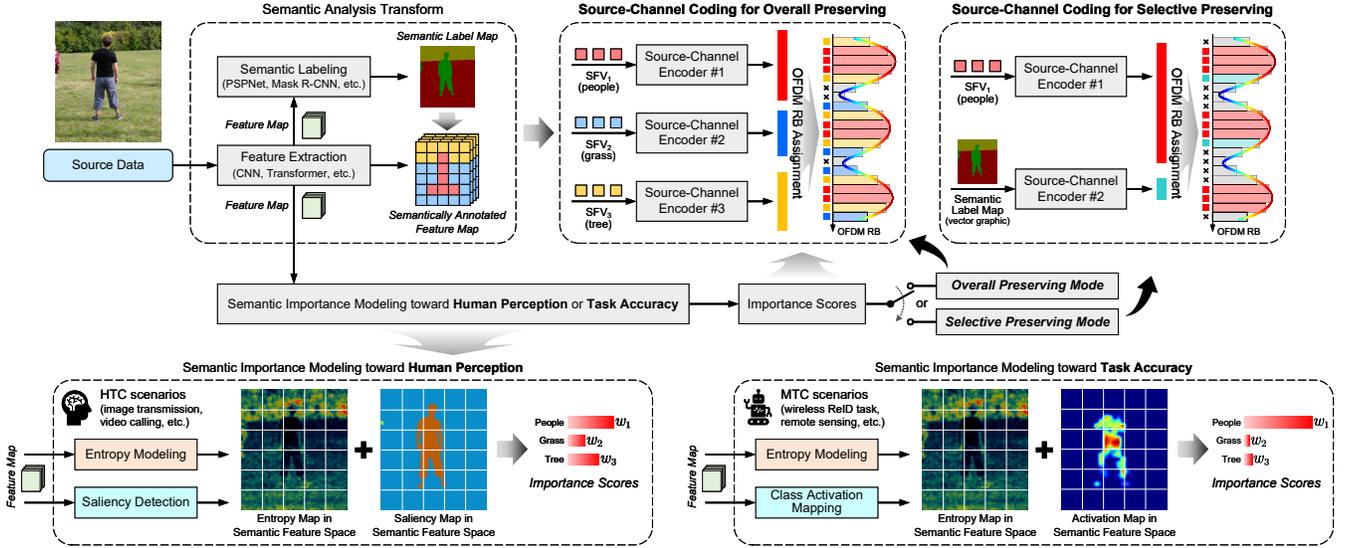}}
	\caption{Detailed procedure of semantic guidance generation and joint source-channel coding, where ``OFDM RB'' denotes the resource block (RB) in orthogonal frequency division multiplexing (OFDM) systems.}
	\label{Fig3}
	\vspace{0em}
\end{figure*}

\subsection{Semantic Guidance Generation}

The technique of \emph{semantic guidance generation} in our SCT embraces two functions, i.e., the semantic analysis transform and the semantic importance modeling. The semantic analysis transform is used for source semantic feature extraction and representation. Different from traditional linear transforms, such as the Karhunen-Lo\`{e}ve transform (KLT), which map the source vector into a code space via a decorrelating invertible transform, the semantic analysis transform maps the source data to the semantic feature space via nonlinear transform operations that can much more closely match and represent the source semantic features \cite{balle2020nonlinear}.

With the right set of parameters, elaborate artificial neural networks (ANNs) with nonlinear properties can approximate arbitrary source distributions and extract the source semantic features. In our SCT system, the ANN-based semantic analysis transform produces the semantically annotated feature map in the semantic feature space. Taking the image source in Fig. \ref{Fig3} as an example, the semantic analysis transform module includes two functions:

\begin{itemize}
  \item \emph{Semantic feature extraction:} It maps the original source to its corresponding feature map at the semantic feature space, which can be usually realized by convolutional neural networks (CNN), Transformer \cite{vaswani2017attention}, etc. The feature map preserves the overall source semantic information that can be used to reconstruct the source data.

  \item \emph{Semantic labeling:} With semantic segmentation networks acting on the feature map, it produces the semantic label map delimiting multiple semantic regions. Combining it with the feature map, one gets the semantically annotated feature map.
\end{itemize}

The other critical function for generating semantic guidance is semantic importance modeling. It first generates the importance maps on the semantic feature space, then the resulting importance scores in terms of every SFV guides the following source-channel coding design. This function is highly tied with the communication purposes \cite{agustsson2019generative}. In particular, we consider two modes of semantic importance modeling, namely
\begin{itemize}
  \item \emph{Importance modeling toward human perception:} For HTC scenarios, as humans can find different regions visually salient in a scene given the context, this mode generates the SFV importance scores combining entropy modeling \cite{balle2020nonlinear} and saliency detection \cite{judd2009learning}.

  \item \emph{Importance modeling toward task accuracy:} For MTC scenarios, as the gradients of task target can directly flow into the neural networks to highlight important regions in the feature map, this mode generates the SFV importance scores by jointly using entropy modeling \cite{balle2020nonlinear} and gradient-weighted class activation mapping (Grad-CAM) \cite{selvaraju2017grad}.
\end{itemize}

We emphasize that for both modes, the entropy modeling is fundamental, which indicates the regional complexity in terms of the source data content. The human subjective perception or task attention is certainly a semantic refinement to the syntax entropy. As depicted in Fig. \ref{Fig3}, the final importance map is a weighted sum over the entropy map and the saliency/activation map, where the specific weighting manner varies with scenarios. The semantic importance score of each SFV is obtained by counting up the importance values of feature map embeddings that belong to this SFV.

The receiver in SCT system invokes the semantic synthesis transform module to reconstruct the source data or to execute downstream tasks directly. As shown in Fig. \ref{Fig2}, the semantic synthesis transform module mainly involves two functions, semantic distortion correction and semantic feature fusion, which are both nonlinear transforms. Explanations about semantic distortion correction will be given later since it plays a critical role in enhancing the fidelity of transmitted data in SCT. As for semantic feature fusion, it actually performs the inverse operation of the transmitter utilizing the corrected feature map to reconstruct the source data.

\subsection{Semantics-Guided Source and Channel Coding}

The aforementioned semantic guidance includes SFVs and their importance scores. It describes the source \emph{data semantics diversity}, i.e., feature map embeddings and data samples are certainly not equally important. Similarity, wireless channels are of the \emph{channel diversity} naturally, e.g., time and frequency selectivity in OFDM, spatial selectivity in MIMO, etc. Shannon theory focuses on how to well utilize the channel diversity to maximize the system capacity, which however discards the data diversity, considering data bits were born equal. For a long time, a question is how to exploit both types of diversity. Our SCT framework provides one solution by injecting semantic guidance into the design of source and channel coding. On the whole, source-channel coding plays a dual-functional role that compresses the input data while also combats the corruptions brought by wireless channels. In particular, as depicted in Fig. \ref{Fig3}, our coding method includes two modes:
\begin{itemize}
  \item \emph{Overall preserving:} This coding mode will be used when the semantic importance scores \emph{differ slightly}. It preserves the overall source content to support the reconstruction of all regions with a high degree of detail at the receiver.

  \item \emph{Selective preserving:} This coding mode will be adopted when the semantic importance scores \emph{differ significantly}. It completely generates parts of the source data from its semantic label map while preserving user-defined regions with a high degree of detail.
\end{itemize}

The overall preserving mode is often used in general HTC scenarios, where the communication purpose wants to preserve the full source data content as well as possible, while falling back to synthesized content instead of some visually-distorted regions for which not sufficient channel bandwidth is available to well transmit corresponding SFVs. The goal is high-fidelity data transmission to please human perception of all the source data contents. The selective preserving mode will be employed in most MTC scenarios and some specified HTC scenarios given the context. For example, in a human video call scenario \cite{agustsson2019generative} where the user wants to fully preserve people in the video stream, but a visually pleasing synthesized background from the semantic label map can serve the purpose as well as the true background. For both modes, our source-channel coding design obeys the following rules:
\begin{itemize}
  \item \emph{Heterogeneous transmitted data formats:} For the overall preserving mode, sequences sent into source-channel encoders are all these SFVs. But for the selective preserving mode, our SCT system jointly transmits SFVs to reconstruct regions of interest and the semantic label map as tiny side information to synthesize other contents.

  \item \emph{Differentiated coding rates:} For SFVs with higher semantic importance scores, they will be allocated more coded symbols, and vice versa.

  \item \emph{Matched channel resource assignment:} For more important SFVs, their source-channel coded symbols will be transmitted over better conditioned channels, e.g., OFDM resource blocks (RBs) with higher channel gain.
\end{itemize}

Herein, source-channel coding can be implemented in either a modular way as that in traditional systems or an integrated way with one module, which will be introduced later. The first design rule indicates a heterogeneous coding formats under the selective preserving mode. Apart from the semantic critical regions transmitted by SFVs, other parts are alternated as the semantic label map as depicted in Fig. \ref{Fig3}, which is formulated as a vector graphic. This amounts to a quite small and source dimension independent overhead in terms of source-channel coding cost. It contributes to a considerable overall reduction in channel bandwidth cost. In the receiver, after source-channel decoding, our SCT system seamlessly combines the preserved content which is generated from SFVs with the synthesized content which is generated from the semantic label map. This method can faithfully preserve the whole source semantics for MTC and some specified HTC scenarios.

As for the second and third rules, they are jointly considered to determine how many and which OFDM RBs are assigned to transmit source-channel coded symbols of each SFV. These RBs with higher channel gain will be of priority to be assigned to more important SFVs as shown in Fig. \ref{Fig3}. In addition, the number of RBs assigned to an SFV is counted by aligning the RB sum channel capacity with the importance score of this SFV. For various scenarios, since the SFV importance scores are of different computation manners as aforementioned, the importance score is certainly of different meaning. Thus, the specific approach to map the importance score with the source-channel coding rate and channel resource assignment will be a critical issue to be optimized individually.

\subsection{Semantic Distortion Correction}

Following the aforementioned two techniques, the third key technique to improve the performance SCT system is \emph{semantic distortion correction}. As shown in Fig. \ref{Fig2}, it can exploit both intra- and inter-SFV correlations to further correct the residual errors left by source-channel decoding. This technique can be indeed categorized as one application of solutions for \emph{inverse problems} in deep learning \cite{ongie2020deep}. In other words, it operates on the representation in semantic feature space to recover the data in source space. As we should note, in our SCT system, semantic distortion correction is highly tied with previous two techniques. In particular, more efforts of distortion correction will be put on these less important SFVs, which often incurs more left errors due to the less allocated channel transmission resources. Fortunately, they are usually easy to be restored due to the strong side information provided by the well transmitted SFVs of higher importance. This process is quite different from traditional post-processing tools used to correct residual errors after source-channel decoding. In that case, the residual error distribution is not relevant to the source semantics. Thus, it will be inefficient and ineffective to restore these distortions due to the lack of semantic guidance generation and the biased coding strategy.

\begin{figure}[t]
	\setlength{\abovecaptionskip}{0.cm}
	\setlength{\belowcaptionskip}{-0.cm}
	\centering{\includegraphics[scale=0.34]{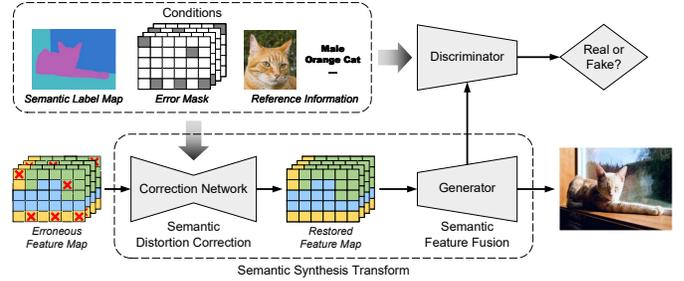}}
	\caption{Detailed procedures of semantic distortion correction.}
	\label{Fig4}
	\vspace{0em}
\end{figure}

Specifically, the semantic distortion correction technique can be realized by generative neural networks, such as the generative adversarial network (GAN), which are of ``creativity'' by exploiting the semantic-level correlations to generate distortion parts of the source content. These deep learning methods have been studied widely in solving inverse problems, such as image post-editing, inpainting, denoising \cite{inpainting}, etc. Herein, it is a new application in communications. A detailed procedure is demonstrated in Fig. \ref{Fig4}, where the goal of SCT is to reconstruct image. The overall structure exploits conditional GANs which introduce additional information into GANs as a conditional model. This additional information can be the semantic label map, error masks, local reference images, or even class labels, etc. The distortion correction network restores the erroneous feature map according to these conditions and then feeds it to the generator (semantic feature fusion module) to reconstruct the source image.

We emphasize that the semantic distortion correction technique often needs some additional side information as conditions. Partial side information needs to be transmitted, such as the semantic label map formulated as a simple vector graphic \cite{agustsson2019generative}, while some other side information like references can be stored locally as knowledge base. However, the coded transmission costs of communicated side information are usually marginal compared to that in the primary data link to transmit necessary SFVs. This almost-negligible overhead exchanges a considerable system performance improvement, which will finally contributes to the reduction in the number of automatic repeat requests (ARQs) that leads to a higher link throughput and lower latency. On the whole, it is worth to transmit the marginal side information supporting the semantic distortion correction.

\section{Implementations}

\begin{figure*}[t]
	\setlength{\abovecaptionskip}{0.cm}
	\setlength{\belowcaptionskip}{-0.cm}
	\centering{\includegraphics[scale=0.16]{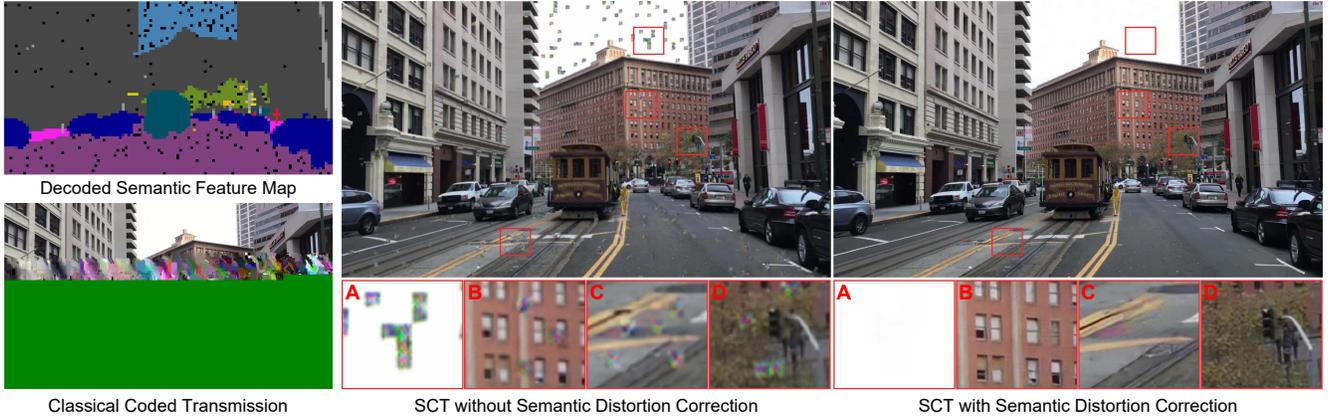}}
	\caption{A comparison of image transmission via two coded transmission schemes. The semantic distortion correction module applied in the modular SCT implementation can enhance the final quality of reconstructed image by utilizing context and classification information.}\label{Fig5}
	\vspace{0em}
\end{figure*}

Practical implementations of source-channel coding fall into two categories, depending on whether the signals transmitted over communication channels are digital or analog. Given the semantic guidance, source and channel coding can either resort to classical coding schemes combined with digital modulation to produce digital signals, or adopt the emerging joint source-channel coding schemes operated by deep neural networks to produce analog signals directly. According to their architecture, we briefly summarize these two SCT implementations as modular and integrated designs, respectively.

\subsection{Modular Implementation}

As a compatible scheme with current digital communications systems (e.g., 4G and 5G), the modular implementation of SCT retains channel coding and digital modulation modules to transmit digital signals.

As an example, consider an image source in Fig. \ref{Fig2}, the semantic guidance generation module produces the semantically annotated feature map first, which is then divided into multiple SFVs. At the step of source-channel encoding, every SFV is first quantized and interleaved, then compressed by an entropy encoder into bits, and finally loaded into packages. Advanced channel coding, such as low-density parity-check (LDPC) or polar codes, follows to protect the source-coded bits against channel corruptions. The overall source-channel coding rate of each SFV is assigned according to its semantic importance score. In the receiver, the digital demodulation operates on the received signal first, it outputs the soft estimations of digit bits which are sent into their corresponding channel decoder. The channel decoded messages are then fed into the source decoder to reconstruct SFVs. A major disadvantage of this approach is that residual errors in channel decoded messages will incur severe error propagation in entropy decoding. However, in our SCT system, this undesirable phenomenon can be alleviated by the semantic distortion correction module. By generating the corrupted regions according to semantic description and context information, the end-to-end transmission quality can be improved.

Fig. \ref{Fig5} demonstrates an example of modular implementation of SCT for image transmission. Here, a 2048 (height) $\times$ 1024 (width) cityscapes image \cite{cityscapes} is transmitted to the receiver over the additive white Gaussian noise (AWGN) channel with signal-to-noise ratio (SNR) 1dB. Among multiple semantic regions in the cityscapes, humans, vehicles, and buildings are assessed of higher importance, but other natural environments, e.g., sky and trees, are of lower importance. For the classical coded transmission, we plot the reconstructed image of the concatenation of BPG codec followed by an LDPC channel codec. As we can see, it fails to reconstruct all regions due to the residual bit errors left by the channel decoder propagating among other coded bits. For our SCT system, residual channel decoded bit errors also lead to a corrupted semantic feature map (plotted in the left top, where the black pixels indicate the corrupted locations), then cause the mosaics during the reconstruction process. The main difference between classical coded transmission and SCT is that our semantics guides the source-channel coding design to match with the importance score of each SFV. In this way, errors tend to cause corruption in SFVs with lower importance. As shown in the right sub-figure, such corruptions can be further corrected by using an inpainting model \cite{inpainting} so that the image quality increases visibly.

\subsection{Integrated Implementation}

Following the joint source-channel coding idea, the source-channel coding part in the SCT system can also be designed as an integrated module, which does not rely on explicit codes for either compression or error correction; Instead, it maps SFVs to channel input analog symbols directly. The encoder and decoder functions are parameterized as two jointly trained ANNs. They together constitute an autoencoder affected by a non-trainable layer that represents the wireless communication channel \cite{djscc}. The joint coding rate of each SFV is aligned with its semantic importance score. The learned source-channel coding allows gradients propagating between coding modules and semantics guidance generation modules. This integrated implementation manner can thus achieve end-to-end learning to minimize the perceptual loss or handle the downstream tasks directly. The whole procedure is also depicted in Fig. \ref{Fig2}.

\begin{figure}[t]
	\setlength{\abovecaptionskip}{0.cm}
	\setlength{\belowcaptionskip}{-0.cm}
	\centering{\includegraphics[scale=0.62]{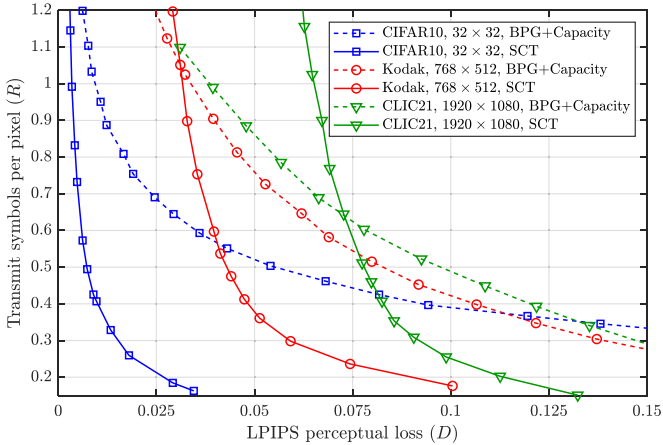}}
	\caption{The quantitative comparison of transmission rate-distortion (RD) performance over the AWGN channel at SNR = 1dB, where three datasets with different resolutions are adopted.}
	\label{Fig6}
	\vspace{0em}
\end{figure}

We exemplify the SCT performance based on the integrated implementation and analog transmitted signals. For a general HTC scenario where images are transmitted over wireless links to meet human browsing requirements, to quantify the effectiveness of SCT, we adopt the emerging learned perceptual image patch similarity (LPIPS) metric to indicate the human perception \cite{LPIPS}. Fig. \ref{Fig6} presents the quantitative comparison of transmission rate-distortion (RD) performance over three datasets of different resolutions. The distortion $D$ is quantified as the LPIPS loss (lower is better), and the transmission rate is defined as the channel bandwidth ratio $R$ (defined as the number of transmitted symbols per pixel). As comparison, we provide the performance of BPG compression combined with the ideal capacity-achieving channel code family to serve as a bound of the traditional coded transmission \cite{djscc}. We find that the integrated SCT can well surpass the traditional scheme especially in the low rate region ($R < 0.6$). In the high rate region, traditional coding performs better due to its elaborate design. Even so, as an emerging scheme, our SCT has shown its potential, which can be further improved by using more advanced neural network architectures and elaborate designs.

\section{Challenges and Future Research}

We identify the main challenges in both theoretical and technical aspects, and provide recommendations in each direction.

\subsection{Theoretical Aspects}

In contrast to traditional coded transmission under the umbrella of Shannon information theory, the SCT system based on deep learning methods lack solid mathematical foundations for theoretical analysis. Consequently, errors may not be well-bounded when the semantic nonlinear transform and semantic distortion correction techniques (such as GANs) are used. Due to this, the transmission rate of SCT may not be well-bounded either, nevertheless, this is desired to guide the practical design of SCT and gain more insights into the performance limits.

A potential solution to address this challenge is expanding the concept of typical sequences in the Shannon information theory with respect to the semantic aspects. By performing reasonable merging operations on the typical sequences belonging to the same semantics, one may try to analyze the asymptotic performance of SCT. In addition, the Kolmogorov complexity, defined as the shortest computer program describing the source sequence, may be used to quantify the rate of semantic compression and analyze the tradeoff function between the channel transmission rate and the end-to-end semantic distortion, i.e., the semantic-level RD function.

\subsection{Technical Aspects}

\subsubsection{Semantic Guidance Generation}

Semantic nonlinear transform plays a key role in the semantic feature extraction and fusion, which relies on deep neural networks as the universal function approximators and stochastic optimization of the end-to-end RD performance. In the SCT system, functions in the semantic guidance generation module dominates the whole system performance, it is thus a key challenge to find unified mathematical foundations to guide their design including ANN model selection and optimization. Furthermore, different from the adoptions of nonlinear transform in CV and NLP communities which operate only on source compression problems, herein, as a communication system, the design of nonlinear transform functions need to take the characteristics of wireless channel and residual errors left by source-channel decoding into consideration.

\subsubsection{Semantics-Guided Source and Channel Coding}

We are clear that the bandwidth cost for transmitting each SFV is proportional to its semantic importance score, but their specific quantitative relationship is still pending to be optimized. As we shall note, the quantitative relationship will be heavily tied with the downstream task of SCT, and the precise theoretical guarantees for finding out an optimal solution for rate and channel resource allocation among SFVs remain unsolved or even completely open. Also, the generalization of an optimized rate and resource allocation strategy for different scenarios is somewhat limited. A tradeoff between optimization and generalization needs to be explored.

\subsubsection{Semantic Distortion Correction}

As a post-processing tool that is used to correct residual errors on the semantic feature map, semantic distortion correction techniques usually involve priori information as the local knowledge. Nevertheless, how to obtain precise priori information is the focus of further research, such as the accurate error mask in the feature map, etc. That calls for a collaborate design with source and channel coding functions.

\section{Conclusions}

In this article, we have introduced a novel unified framework of semantics-guided source and channel coding. The corresponding end-to-end communication system is collected under the name semantic coded transmission (SCT). The transmitted content meaning and downstream tasks have been explicitly taken into account for boosting the performance of coded transmission system. In particular, we have presented the SCT architecture, key methodologies, implementation schemes, and performance gains. Also, we have pointed out some challenges and research opportunities on this emerging topic.

\section*{Acknowledgements}

This work was supported in part by the National Natural Science Foundation of China under Grant 92067202, Grant 62001049, Grant 62071058, and Grant 61971062, in part by the Beijing Natural Science Foundation under Grant 4222012, and in part by the Major Key Project of PCL under Grant PCL2021A15.

\ifCLASSOPTIONcaptionsoff
  \newpage
\fi

\bibliographystyle{IEEEtran}

\bibliography{Ref}

\end{document}